\documentstyle[prl,aps,epsf]{revtex}
\begin{document}
\title{Finite-size scaling for the S=1/2 Heisenberg 
Antiferromagnetic Chain}
\author{Victor Barzykin$^{*,**}$, Ian 
Affleck$^{**,\dagger}$}
\address{$^*$National High Magnetic Field Laboratory, 
Florida State University,\\
1800 E. Paul Dirac Dr., Tallahassee, Florida 32306-4005 
\\
and \\
$^{**}$Department of Physics and Astronomy and 
$^{\dagger}$Canadian Institute for
Advanced Research, \\ University of British Columbia, 
Vancouver, BC,
Canada, V6T 1Z1}
\maketitle \begin{abstract}
Corrections to the asymptotic correlation function in a Heisenberg spin-$1/2$ 
antiferromagnetic spin chain are known 
to vanish slowly (logarithmically) as a function of the distance $r$ or 
the chain size $L$. 
This leads to significant differences with 
numerical results. We calculate the sub-
leading logarithmic corrections
to the finite-size correlation function, using 
renormalization group improved
perturbation theory, and compare the result with numerical data.

\end{abstract}

The correlation function in the spin-$1/2$ Heisenberg 
antiferromagnetic
chain is difficult to determine from the Bethe Ansatz, 
so other
methods are used for this purpose, such as bosonization
and conformal field theory(CFT). These methods work very
well as a tool to determine long-distance asymptotics.
Numerical work\cite{Kubo,Liang,Sandvik,Hallberg,Koma}, 
however, often questioned this approach. This has
been explained by the fact that finite size corrections 
vanish slowly at long distances, as $1/\ln{L}$, and 
$1/\ln (r)$, where 
$L$ is the system size (periodic boundary conditions 
assumed) and $r$
the separation of the 2 points, due 
to the presence of a marginally irrelevant 
operator\cite{Affleck,Singh}.

Following the methods and notation of 
\cite{Affleck,Affleck1},the continuum limit of the 
Heisenberg model can be written in $SU(2)$ - symmetric
form using non-Abelian bosonization\cite{Affleck}. The 
action for the $SU(2)$-symmetric
matrix field ${\bf g}^{\alpha}_{\beta}$ includes the Wess-
Zumino term with coefficient
$k=1$. This theory is equivalent to a free boson defined 
on a circle.
The low-energy Hamiltonian in the continuum 
approximation is given by:
\begin{equation}
\label{ham}
H = H_0 - 8 \pi^2/\sqrt{3} \lambda {\bf J}_L \cdot {\bf 
J}_R,
\label{hamilt}\end{equation}
where $H_0$ is the Hamiltonian density for a free boson, 
${\bf J}_{L,R}$ 
are left and right $SU(2)$ currents:
\begin{equation}
{\bf J}_L \equiv { -i \over 4 \sqrt{\pi}} tr[{\bf g}^{\dagger} 
\partial_-{\bf g}\bbox{\sigma}], \ \ \
{\bf J}_R \equiv { i \over 4 \sqrt{\pi}} tr[\partial_+{\bf g} 
{\bf g}^{\dagger}\bbox{\sigma}]
\end{equation}
(Here we use the notation $\lambda$ for the marginal 
coupling constant
rather than $g$ as in \cite{Affleck1}.)  

We now turn to the discussion of the asymptotic 
correlation function.
The spin operators can be written in non-abelian 
bosonization notation as:
\begin{equation}
{\bf S}_j  = ({\bf J}_L + {\bf J}_R) + \hbox{const}\ i 
(-1)^j tr[{\bf g} \bbox{\sigma}]
\end{equation}
Thus the correlation function has uniform and staggered 
terms,
\begin{equation}
G(r)=\langle S_0^z S_r^z \rangle \rightarrow G_u(r) + (-
1)^r G_s(r).
\end{equation}
Both terms vary slowly on the scale of the lattice 
spacing, and
correspond to different Green's functions in the 
continuum theory
In this paper we only consider the staggered term 
$G_s(r)$,
\begin{equation}
\label{stc}
G_s(r) \propto \langle tr(\bbox{\sigma}^z {\bf g})(r) tr(\bbox{\sigma}^z 
{\bf g})(0)\rangle_L.
\end{equation}

The staggered correlation function for a finite chain 
obeys
the following renormalization group (RG) equations\cite{Affleck}:
\begin{equation}
\label{RG}
[\partial/\partial \ln{r} 
+\beta(\lambda)\partial/\partial \lambda + 
2 \gamma(\lambda)] G_s(r,r/L,\lambda) = 0,
\end{equation}
where $\beta(\lambda)$ is the beta function for the 
coupling constant
$\lambda$ in Eq.(\ref{hamilt}):
\begin{equation}
{d \lambda \over d \ln{r}} = \beta(\lambda) 
\label{beta}
\end{equation}
and $\gamma(\lambda)$ is the anomalous dimension.  In Eq. (\ref{RG})
the $r$-derivative acts only on the first argument of $G_s$; $r/L$ is
held fixed.  Eq. (\ref{RG}) expresses
the fact that a rescaling of both lengths $L$ and $r$ by 
a common factor can
be compensated for by a change in the effective coupling 
constant, $\lambda (r)$
and a rescaling of the correlation function.  
The solution of Eq.(\ref{RG}) has the form:
\begin{equation}
\label{sol}
G_s(r,\lambda_0) = exp\left(-2 \int_{r_0}^r d \ln{r'} 
\gamma[\lambda(r')]\right)
F[r/L,\lambda(r)],
\end{equation}
where $\lambda_0 \equiv \lambda(r_0)$ is the "bare" 
coupling - a coupling at the
ultraviolet cutoff scale $r_0$, $F[r/L,\lambda(r)]$ is 
an arbitrary function of the
effective coupling constant at scale $r$, $\lambda(r)$.

The coupling constant flows to zero as the distance $r$ 
is increased, and
one can use perturbative expressions for 
$\gamma(\lambda)$ and 
$\beta(\lambda)$ to determine long-distance properties. 
The universal
terms in the perturbative expansion for the $\beta$-
function\cite{Solyom}
and the anomalous dimension\cite{Affleck,Singh} are 
known,
\begin{eqnarray}
\label{bta}
\beta(\lambda) &=& - (4 \pi/\sqrt{3}) \lambda^2 - 
(1/2)(4 \pi/\sqrt{3})^2 
\lambda^3 \\
\label{gmma}
\gamma(\lambda) &=& 1/2 - (\pi/\sqrt{3}) \lambda.
\end{eqnarray}
Thus the effective coupling is given by:
\begin{equation} {1\over \lambda (r)}-{1\over \lambda
_0}=(4\pi /\sqrt{3})\{\ln (r/r_0)+(1/2 )\ln [\ln 
(r/r_0)]\}+ O(1).\label{lambda(r)}
\end{equation}
Rewriting the integral in Eq.(\ref{sol}) using 
Eq.(\ref{bta}), 
one easily finds\cite{Affleck1}:
\begin{equation}
\label{factor}
\int_{\lambda_0}^{\lambda(r)} 
[\gamma(\lambda)/\beta(\lambda)]
d\lambda = (1/2) \ln(r/r_0) + 
(1/4)\ln[\lambda(r)/\lambda_0] + \dots
\end{equation}
Thus the Green's function has the expansion:
\begin{equation}
G_s(r,L,\lambda )={1\over r}\sqrt{\lambda_0\over \lambda 
(r)}e^{\sum_{n=1}^\infty a_n[\lambda (r)^n-
\lambda_0^n]}\sum_{m=0}^\infty F_m(r/L)\lambda (r)^m 
\label{GF}\end{equation}
The coefficients, $a_n$ and the functions $F_m(r/L)$ can 
be determined by
doing perturbation theory in the bare coupling constant 
and then recasting the
resulting expression in terms of the renormalized 
coupling constant in the form
of Eq. (\ref{GF}).  This automatically incorporates, at 
low orders, the leading
log divergences to all orders in perturbation theory.  
This method is standard in
Quantum Chromodynamics calculations and is known as 
``renormalization group improved
perturbation theory''. 

The zeroth order term of this sum, $F_0(r/L)$, is given 
by the free theory - the conformally
invariant WZW model on a circle of length L, and can be 
obtained by
conformal transformation. Since ${\bf g}$ has scaling 
dimension 1/2, for an infinite system:
\begin{equation}
<tr [{\bf g}(r)\bbox{\sigma}^z(r)]tr [{\bf g}(0)\bbox{\sigma}^z]>={1\over 
r}.\label{gcorr}\end{equation}
(Here we have chosen a convenient normalization for the 
operator ${\bf g}$.)  For a finite
system with periodic boundary conditions, we make a 
conformal transformation from the
infinite plane to the cylinder, obtaining:
\begin{equation}
<tr [{\bf g}(r)\bbox{\sigma}^z]tr [{\bf g}(0)\bbox{\sigma}^z]>_L
={\pi \over L\sin (\pi r/L)}.\end{equation}
Thus we see that:
\begin{equation} F_0(r/L)\propto {(\pi r/L)\over \sin 
(\pi r/L)}.\end{equation}
Using Eq. (\ref{lambda(r)}), we thus obtain the 
asymptotic correlation function:
\begin{equation}G_s(r,L)\to {A \over (\pi /L)\sin (\pi 
r/L)}\{\ln (r/r_0)+(1/2)\ln [\ln (r/r_0)]\}.
\end{equation}
Essentially this result was obtained in \cite{Affleck1}. 
For related work on the correlation function for the infinite length spin
chain see Ref. \onlinecite{Lukyanov}. 

The new result which
we derive here is the next order correction, $F_1(r/L)$, 
in Eq. (\ref{GF}).  The
first order perturbation theory result, using the 
Hamiltonian of Eq. (\ref{ham}),
gives:
\begin{equation}
\delta <tr [{\bf g}(r)\bbox{\sigma}^z(r)]tr 
[{\bf g}(0)\bbox{\sigma}^z]>={8\pi^2\lambda_0 \over \sqrt{3}}
\int d\tau dx{\cal T}<tr [{\bf g}(r,0)\bbox{\sigma}^z]tr 
[{\bf g}(0,0)\bbox{\sigma}^z]{\bf J}_L(x,\tau )\cdot 
{\bf J}_R(x,\tau )>_L,\end{equation}
where $\cal {T}$ denotes time-ordering.
This correlation function can be evaluated using 
standard CFT techniques.  We first
obtain its value for an infinite system, then obtain the 
result for finite $L$ by
conformal transformation.  Using the general result for 
3-point functions of
primary operators we obtain:
\begin{equation}
{\cal T}<tr [{\bf g}(r,0)\bbox{\sigma}^z]tr [{\bf g}(0,0)\bbox{\sigma}^z]{\bf 
J}_L(x,\tau )\cdot 
{\bf J}_R(x,\tau )>\propto {r\over (x+i\tau )(x-r+i\tau 
)(x-i\tau )(x-r-i\tau )}
\end{equation}
The normalization constant can be fixed from the 
operator product expansion (OPE):
\begin{equation}
{\bf J}_L(x,\tau )\cdot {\bf J}_R(x,\tau )tr 
[{\bf g}(0)\bbox{\sigma}^z]\to 
-{tr [{\bf g}(0)\bbox{\sigma}^z]\over 
16\pi^2(\tau^2+x^2)},\end{equation}
and the normalization of the zeroth order Green's 
function in Eq. (\ref{gcorr}).
This gives:
\begin{eqnarray}
&&{\cal T}<tr [{\bf g}(r,0)\bbox{\sigma}^z]tr [{\bf g}(0,0)\bbox{\sigma}^z]{\bf 
J}_L(x,\tau )\cdot 
{\bf J}_R(x,\tau )>\nonumber \\
&&=-{r\over 16\pi^2(x+i\tau )(x-r+i\tau )(x-i\tau )(x-r-
i\tau )}.
\end{eqnarray}
The correlation function on the cylinder can be obtained 
by conformal
transformation:
\begin{eqnarray}
&&{\cal T}<tr [{\bf g}(r,0)\bbox{\sigma}^z]tr [{\bf g}(0,0)\bbox{\sigma}^z]{\bf 
J}_L(x,\tau )\cdot 
{\bf J}_R(x,\tau )>_L\nonumber \\
&&=-{(\pi /L)^3\sin (\pi r/L)\over 16\pi^2\sin [\pi 
(x+i\tau )/L]\sin [\pi (x-r+i\tau )/L]
\sin [\pi (x-i\tau )/L]\sin [\pi (x-r-i\tau )/L]}.
\end{eqnarray}
Thus the first order correction  is given by:
\begin{eqnarray}
\delta <tr [{\bf g}(r)\bbox{\sigma}^z(r)]tr [{\bf g}(0)\bbox{\sigma}^z]>_L=
 \left( {-\lambda_0\over 2\sqrt{3}}\right) 
\int_{-\infty}^\infty d\tau \int_{0}^L dx 
\nonumber \\
{(\pi /L)^3\sin (\pi r/L)\over 16\pi^2\sin [\pi (x+i\tau 
)/L]\sin [\pi (x-r+i\tau )/L]
\sin [\pi (x-i\tau )/L]\sin [\pi (x-r-i\tau )/L]}.
\end{eqnarray}
This integral has a logarithmic ultraviolet divergence 
at $\tau \to 0$, and $x\to 0$ or $r$.
This can be cut off by restricting the integral to 
$|x|^2+\tau^2>r_0^2$ and $|x-r|^2+\tau^2>r_0^2$.
The resulting logarithmic dependence on $r_0$ is exactly 
what is needed in order
for the resulting expression to have the form of Eq. 
(\ref{GF}), since:
\begin{equation}
\sqrt{\lambda_0\over \lambda (r)}\approx 1+{2\pi 
\lambda_0 \over \sqrt{3}}\ln (r/r_0).
\end{equation}
Remarkably, the integrals can be done exactly.  The $x$-
integral can conveniently
be done first using contour methods.  This gives:
\begin{eqnarray}
&& \delta <tr [{\bf g}(r)\bbox{\sigma}^z(r)]tr 
[{\bf g}(0)\bbox{\sigma}^z]>_L\nonumber \\
&&={i\pi^2\lambda_0\over 2\sqrt{3}L}\int_{-
\infty}^\infty du\left[{1\over \sinh u 
\sinh (u+i\pi r/L)}-\hbox{complex 
conjugate}\right].\end{eqnarray}
(Here $u\equiv 2\pi \tau /L$ and the integral is cut off 
at $|u|>r_0/L$.)
This indefinite integral may be done exactly by changing 
variables to $\tanh u$, giving:
\begin{eqnarray}
&& <tr [{\bf g}(r)\bbox{\sigma}^z(r)]tr [{\bf g}(0)\bbox{\sigma}^z]>_L\nonumber \\
&&={1\over (L/\pi )\sin [\pi r/L]}\{1+{2\pi 
\lambda_0\over \sqrt{3}}
\{\ln [(L/r_0)\sin (\pi r/L)]+\hbox{constant}\}\}
.\label{GF1}\end{eqnarray}
This is the result of perturbation theory to first order 
in the bare coupling
constant $\lambda_0$.  The next step is to 
``renormalization group improve''
this result by matching it to the expression in Eq. 
(\ref{GF}). Expanding this
expression to first order in the bare coupling constant, 
using Eq. (\ref{lambda(r)})
gives:
\begin{equation}
G_s(r)\propto {1\over (L/\pi )\sin [\pi 
r/L]}[1+\lambda_0(2\pi /\sqrt{3})\ln (r/r_0)]
[1+\lambda_0F_1(r/L)/F_0(r/L)].\label{GF2}\end{equation}
Comparing Eq. (\ref{GF1}) and (\ref{GF2}) we see that:
\begin{equation}
F_1(r/L)/F_0(r/L)={2\pi \over \sqrt{3}}\ln \left[{L\over 
r}\sin 
\left({\pi r\over 
L}\right)\right]+\hbox{constant}.\end{equation}
Thus our renormalization group improved expression for 
the correlation function is:
\begin{equation}
G_s(r,L,\lambda_0)\propto {1\over (L/\pi )\sin [\pi 
r/L]}\sqrt{\lambda_0\over 
\lambda (r)}\left\{1+\lambda (r)\left\{{2\pi \over 
\sqrt{3}}\ln \left[{L\over r}\sin 
\left({\pi r\over 
L}\right)\right]+\hbox{constant}\right\}\right\}.
\label{GF3}\end{equation}
The advantage of this RG improved expression is that we 
may now go to arbitarily
large $r$, a limit in which the large logarithms, $\ln 
(r/r_0)$ invalidate
finite-order perturbation theory and infinite 
resummations of most divergent
diagrams are neccessary.  This is automatically taken 
care of by Eq. (\ref{GF3})
together with the expression for $\lambda (r)$ in Eq. 
(\ref{lambda(r)}).  In
this asymptotic limit we may use:
\begin{equation}
\label{lmbd}
\lambda (r)\approx {\sqrt{3}\over 4\pi 
\ln (r/r_0)},
\end{equation}  
for the factor of $\lambda (r)$ inside the curly 
brackets in Eq. (\ref{GF3}).
This gives:
\begin{eqnarray} G_s(r)&\to& {A\over (L/\pi )\sin [\pi 
r/L]}
[\ln (Cr/r_0)+(1/2)\ln [\ln (r/r_0)]]^{1/2}\nonumber \\
&&\cdot \left\{1+{1\over 2\ln (r/r_0)}\left\{\ln 
\left[{L\over r}\sin 
\left({\pi r\over 
L}\right)\right]+\hbox{constant}\right\}\right\},
\label{GF4}\end{eqnarray}
where:
\begin{equation} C\approx e^{\sqrt{3}/4\pi 
\lambda_0+O(1)}.\end{equation}
We now see that the ``constant'' term can be adsorbed, 
to lowest order in $1/\ln (r/r_0)$
, into a rescaling of $C$ (i.e. a shift of $\lambda_0$) 
so we henceforth drop it.  This is all
information that can be extracted from the RG to this 
order.  Two 
non-universal free parameters
remain: the overall amplitude, $A$ and the bare coupling 
$\lambda_0$, appearing
as the constant $C$ in Eq. (\ref{GF4}).  The amplitude 
$A$ was recently
determined, from Bethe ansatz results\cite{Lukyanov} for the S=1/2 
Heisenberg model, to be 
\cite{Affleck1}
$A=(2\pi )^{-3/2}$.  The bare coupling $\lambda_0$ (for 
some conveniently
chosen value of $r_0$) is not known exactly for the 
S=1/2 Heisenberg model.
It can be determined by fitting numerical results.  
Since the same bare coupling
constant also appears in various other finite size 
corrections, including
the finite size spectrum \cite{Affleck} various 
consistency checks could
be made.  However, this requires further calculations to 
ensure consistency
of the cut-off schemes in various calculations and we do 
not attempt it here.  

Note that 
if we chose $z \equiv (L/\pi) \sin(\pi r/L)$ 
instead of $r$ as the length
scale in Eq.(\ref{RG}), we would find, to this
order, that the finite-size correlation function at 
distance $r$ is given by
an infinite chain correlation function at $z$:
\begin{equation}
\label{sss}
G_s(r,r/L) = G_s(z,0)={1\over z (2\pi )^{3/2}}
\left\{\ln (C z/r_0)+(1/2)\ln [\ln (z/r_0)]\right\}^{1/2}.
\end{equation} 
However, this fact does not necessarily hold for higher 
order RG. As follows from our result Eq.(\ref{GF4}),
the finite-size correction to the asymptotic correlation
function vanishes as $r^2/L^2$ as one approaches the infinite-chain 
limit, $r/L \to 0$. The expansion in $1/\ln(r/r_0)$ remains, but
the functions $F_m(r/L)$ (see Eq.(\ref{GF}) approach constants 
in this limit. (We have only checked this for $F_0(r/L)$ and
$F_1(r/L)$)

It is interesting  to compare our result Eq.(\ref{GF4}) 
with phenomenological
expressions used to fit numerical data. Koma and 
Mizukoshi\cite{Koma} used
the scaling function of the form:
\begin{equation}
G_s(r,L) = {A\{\ln[(L/\pi r_0)\sin(\pi r/L)]\}^{1/2} 
\over (L/\pi)\sin(\pi r/L)}
\end{equation}
with  $A = (2
\pi)^{-3/2} \simeq 0.0635$, close to the exact answer.
This form is equivalent to Eq.(\ref{sss}) with the
slow log log term replaced by a constant. 
The best fit was obtained for $A \simeq 0.065$, close to 
the exact answer.
Kubo {\em et al}\cite{Kubo} and Hallberg {\em et 
al}\cite{Hallberg}
defined a scaling function $f(r/L)$,
\begin{equation}
\label{scf}
G_s(r,L) = G_s(r,\infty) f(r/L),
\end{equation}
and adopted a phenomenological expression,
\begin{equation}
\label{phenom}
f(x) \propto [1 + A \sinh^2(B x)]^{2 \eta}.
\end{equation} 
Note that Eq.(\ref{scf}) does not agree with our RG 
analysis. 
They found the best fit for $A=0.28822$, $B=1.673$, $2 
\eta=1.805$.
As it has been noted by one of us, the form is 
remarkably (within $0.05$\verb+%+) 
close to the CFT prediction for the general $xxz$ model,
since
\begin{equation}
1 + 0.28822 \sinh^2(1.673 x) \simeq \left[{\pi x \over 
\sin(\pi x)} \right]^{1/2}
\end{equation}
Further, for $\eta$ close to $1$ we can replace the 
phenomenological formula
Eq.(\ref{phenom}) by an equivalent,
\begin{equation}
[1 + 0.28822 \sinh^2(1.673 x)]^{2 \eta} \simeq
\left[{\pi x \over \sin(\pi x)} \right]\left\{1 + (\eta 
-1) \ln\left[{\pi x \over 
\sin(\pi x)}\right]\right\}.
\end{equation}
The scaling function $f(x)$ is similar to 
$F(\lambda(r),x)$ defined
in Eq.(\ref{GF4}). We find that $\eta$ depends on $r$, 
\begin{equation}
\eta(r) \simeq 1 - {1 \over 2 \ln(c r)}.
\end{equation}

\begin{figure}
\vspace{-3cm}
\hspace{4cm}
\epsfxsize=8cm
\epsfbox{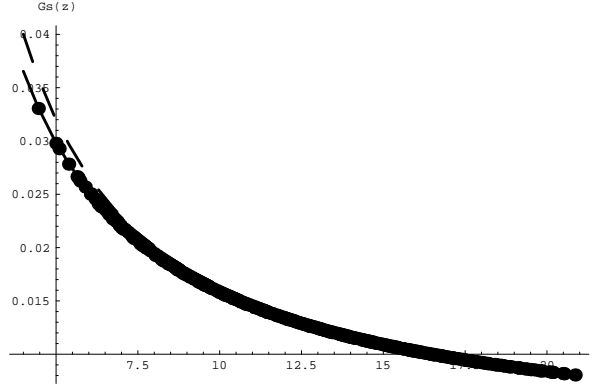}
\vspace{-2cm}
\caption{Scaled DMRG data for spin correlation function
compared with one-loop theoretical expression (solid line)
and free boson result (dashed line)}\label{one}
\end{figure}

\begin{figure}
\vspace{-3cm}
\hspace{4cm}
\epsfxsize=8cm
\epsfbox{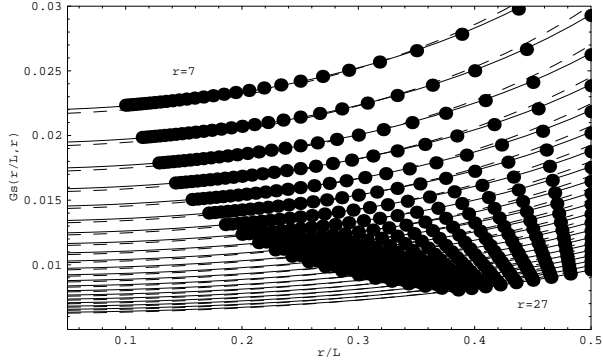}
\vspace{-2cm}
\caption{Numerical data for the spin-spin correlation function
$G_s(r,L)$ from Ref.\protect{\onlinecite{Hallberg}} versus $r/L$ for
$r=7-27$. Solid lines - our one-loop result (with one parameter - $c$),
dashed line - the result without our correction (also with one parameter -
$c$).}\label{two}
\end{figure}

Finally, let us compare our theoretical expression 
Eq.(\ref{GF4}) with DMRG data of Ref.\cite{Hallberg}. 
We have found that the log log term, which is higher
order, is almost constant, and does not influence
the comparison. Since adding a loglog term
introduces one more free parameter,
we decided to drop it. To  
obtain data collapse for $G_s(r,L)$ in
a one-loop RG we use Eq.(\ref{sss}), 
with the log log term dropped and use $C$ as a 
free parameter. 
We compare this with a zero-order non-interating result,
\begin{equation}
\label{free}
G_{free}(z) = const {1 \over z}
\end{equation}
The result is shown in Fig.\ref{one}. Both expressions use one 
free parameter. The better fit produced by our one-loop   
expression is obvious. 
Alternatively, we can use the  result Eq.(\ref{GF4}), 
with $A=1/(2\pi )^{3/2}$,
the log log term dropped, and C taken as a free parameter.  (These
two expressions are the same to the order that we have calculated
in $1/\ln r$.)
The comparison of this formula with numerical data is
shown in Fig.\ref{two}. For comparison we also show the result without the
correction that we have calculated,
\begin{equation} 
G_s(r) = {A [\ln (Cr/r_0)]^{1/2} \over (L/\pi )\sin [\pi r/L]}
\label{GFO}
\end{equation} 
\begin{figure}
\vspace{-3cm}
\hspace{4cm}
\epsfxsize=8cm
\epsfbox{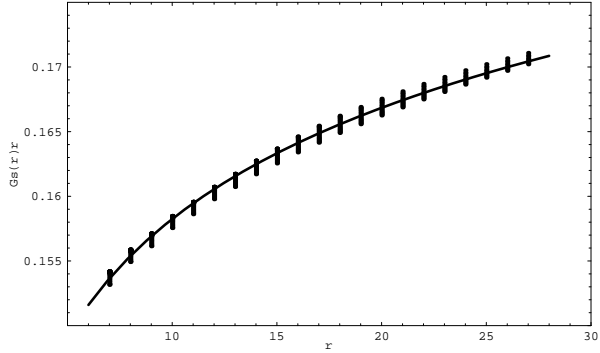}
\vspace{-2cm}
\caption{Spin-spin correlator multiplied by $r$, $r G_s(r,\infty)$,
vs $r$ extracted from the DMRG data of Ref.\protect{\onlinecite{Hallberg}}.
Solid line - the one-loop result.} \label{three}
\end{figure}
                       
It is instructive to find the value of $A$ from the numerical data
to compare it with exact value. 
Using Eq.(\ref{GF4}) with $A$ and $C$ as 
free parameters to fit the numerical data, we find $A = 0.0636427$, 
which differs only $0.2$\verb+%+ from the exact answer. This is much
better than what we would have found without the correction
calculated in this paper, $A = 0.0578896$, which is off by some 
$9$\verb+%+.

Finally, we can extract from
the data the behavior of the correlation function at
$L=\infty$, using Eq.(\ref{GF}) and dividing it by
\begin{equation}
F_0(r/L) + F_1(r/L) \lambda(r) = 
\left\{1 + {1 \over 2\ln(c r) } \ln\left[{L \over \pi r}
\sin\left({\pi r \over L}\right)\right]\right\}{\pi r \over L
\sin[\pi r/L]}
\end{equation}
The result is given by
\begin{equation}
G_s(r,\infty) = {\sqrt{\ln(c r)} \over (2 \pi)^{3/2} r},
\end{equation}
with the same constant $c$. This way of extracting $L=\infty$
behavior from finite-size data is different from the phenomenological
scaling expression that Hallberg {\em et al.} used  in 
Ref.\onlinecite{Hallberg}, $G_s(r,L) = G_s(r,\infty) f(r/L)$.
The result is shown in Fig.\ref{three}.
  
In conclusion, we have shown that bosonization approach
provides an accurate description of the spin correlation
function in a finite spin-$1/2$ Heisenberg spin chain.
Our theoretical result Eq.(\ref{GF4}) compares favorably
with numerical data at long length scales.
We are greatful to K. Hallberg for providing us 
numerical data from
Ref.\cite{Hallberg}. This work was supported by NSERC of 
Canada. One of us (VB) acknowledges support by the NHMFL 
through NSF cooperative agreement No. DMR-9527035 and the
State of Florida.


\begin{references}
\bibitem{Kubo} K. Kubo, T.A. Kaplan and J. Borysowicz, 
Phys. Rev. {\bf B38},
11550 (1988).
\bibitem{Liang} S. Liang, Phys. Rev. Lett. {\bf 64}, 
1597 (1990).
\bibitem{Sandvik} A.W. Sandvik and D.J. Scalapino, Phys. 
Rev. {\bf B47},
12333 (1993).
\bibitem{Hallberg} K. Hallberg, P. Horsch and G. 
Martinez, Phys. Rev. {\bf
B52}, R719 (1995).
\bibitem{Koma}T. Koma and N. Mizukoshi, J. Stat. Phys. 
{\bf 83}, 661 (1996).
\bibitem{Affleck} I. Affleck, D. Gepner, H.J. Schulz and 
T. Ziman, J. Phys.
{\bf A22},
511 (1989).  For a review see I. Affleck, {\it Fields, 
Strings and Critical
Phenomena}
[ed. E. Br\'ezin and J. Zinn-Justin, North-Holland, 
Amsterdam, 1989]; 511.
\bibitem{Singh} R.R. Singh, M.E. Fisher and R. Shankar, 
Phys. Rev. {\bf B39},
2562 (1989).
\bibitem{Affleck1}  I. Affleck, J. Phys. {\bf A31}, 4573 
(1998).
\bibitem{Solyom} J. Solyom, Adv. Phys. {\bf 28}, 201
(1979).
\bibitem{Lukyanov}S. Lukyanov and A. Zamalodchikov, 
Nucl. Phys. {\bf B493}, 571 (1997); 
S. Lukyanov, preprint, cond-mat/9712314.
\end{references}
\end{document}